\documentclass[sn-mathphys-num]{sn-jnl}

\usepackage{graphicx}%
\usepackage{multirow}%
\usepackage{amsmath,amssymb,amsfonts}%
\usepackage{amsthm}%
\usepackage{mathrsfs}%
\usepackage[title]{appendix}%
\usepackage{xcolor}%
\usepackage{textcomp}%
\usepackage{manyfoot}%
\usepackage{booktabs}%
\usepackage{algorithm}%
\usepackage{algorithmicx}%
\usepackage{algpseudocode}%
\usepackage{listings}%

\newtheorem{theorem}{Theorem}
\newtheorem{proposition}[theorem]{Proposition}%
\newtheorem{example}{Example}%
\newtheorem{remark}{Remark}%

\usepackage{tikz}
\usetikzlibrary{positioning}
\usepackage{graphicx}
\usepackage{tikz,pgfplots,pgfplotstable}
\usepgfplotslibrary{fillbetween}
\usetikzlibrary{patterns}
\pgfplotsset{compat=1.16}

\definecolor{mayablue}{rgb}{0.45, 0.76, 0.98}
\definecolor{oceanboatblue}{rgb}{0.0, 0.47, 0.75}
\definecolor{pastelred}{rgb}{1.0, 0.41, 0.38}

\begin{document}

\title[Source-sink dynamics in a two-patch SI epidemic model]{Source-sink dynamics in a two-patch SI epidemic model with life stages and no recovery from infection}


\author[1]{\fnm{Jimmy} \sur{Calvo-Monge}}\email{jimmy.calvo@ucr.ac.cr}
\author*[2]{\fnm{Jorge} \sur{Arroyo-Esquivel}}\email{jarroyoesquivel@carnegiescience.edu}
\author[3]{\fnm{Alyssa} \sur{Gehman}}\email{alyssa.gehman@hakai.org}
\author[1,4]{\fnm{Fabio} \sur{Sanchez}}\email{fabio.sanchez@ucr.ac.cr}

\affil[1]{\orgdiv{Escuela de Matem\'atica}, \orgname{Universidad de Costa Rica}, \orgaddress{\city{San Pedro}, \postcode{11501}, \state{San José}, \country{Costa Rica}}}

\affil*[2]{\orgdiv{Department of Global Ecology}, \orgname{ Carnegie Institution for Science}, \orgaddress{\city{Washington DC}, \postcode{20015}, \country{United States of America}}}

\affil[3]{\orgname{Hakai Institute}, \orgaddress{\postcode{25039}, \state{British Columbia}, \country{Canada}}}

\affil[4]{\orgdiv{Centro de Investigaci\'on en Matem\'atica Pura y Aplicada}, \orgname{ Universidad de Costa Rica}, \orgaddress{\city{San Pedro}, \postcode{11501}, \state{San José}, \country{Costa Rica}}}



\abstract{This study presents a comprehensive analysis of a two-patch, two-life stage SI model without recovery from infection, focusing on the dynamics of disease spread and host population viability in natural populations. The model, inspired by real-world ecological crises like the decline of amphibian populations due to chytridiomycosis and sea star populations due to Sea Star Wasting Disease, aims to understand the conditions under which a sink host population can present ecological rescue from a healthier, source population. Mathematical and numerical analyses reveal the critical roles of the basic reproductive numbers of the source and sink populations, the maturation rate, and the dispersal rate of juveniles in determining population outcomes. The study identifies conditions for disease-free, endemic, and extinction equilibria in sink populations, emphasizing the potential for ecological and evolutionary mechanisms to facilitate coexistence or recovery. These findings provide insights into managing natural populations affected by disease, with implications for conservation strategies, such as the importance of maintaining reproductively viable refuge populations and considering the effects of dispersal and maturation rates on population recovery. The research underscores the complexity of host-pathogen dynamics in spatially structured environments and highlights the need for multi-faceted approaches to biodiversity conservation in the face of emerging diseases.}



\maketitle

\section{Introduction}
Many natural populations are at risk of extinction due to disease-driven mortality, making managing these diseases a crucial endeavor to protect these species \cite{McCallum2012-si}. One of the most documented cases of this population decline is the mortality of amphibian populations caused by chytridiomycosis, a disease caused by the fungal pathogen \textit{Batrachochytrium dendrobatidis} (Bd) \cite{Li2021-om}. More recently, sea star populations have been declining due to the presence of an unidentified pathogen that causes the Sea Star Wasting Disease (SSWD) \cite{Miner2018-vc}. Management of diseases in natural populations is a task that requires understanding both the transmission dynamics of such diseases as well as the population dynamics of the species studied. Ecological models that consider the role of the transmission of infectious diseases have been studied for almost half a century \cite{Anderson1979-rj,May1979-gk}. Host population decline can be caused when recovery from the disease is significantly slower than disease-induced mortality \cite{Palomar2023-du} or when recovery events are rare \cite{Murray2009-ev}. However, classical models undermine the possibility of host populations going extinct, observed empirically and likely caused by demographic stochasticity at low populations or the presence of pathogen reservoirs in the environment \cite{De_Castro2005-ra}.

Host population decline caused by disease-induced mortality has been observed to vary spatially. For example, a lower transmission rate of Bd in bromeliad microhabitats leads to higher recruitment of amphibian juveniles than in stream microhabitats \cite{Blooi2017-qo}. In the case of SSWD, higher water temperatures in some regions potentially hinder the immune response of the sea stars to the disease, causing increased disease-induced mortality \cite{Aalto2020-mu}. This spatial variation begs how disease transmission is affected when individuals can disperse between patches. The role of mobility in disease modeling has been recently identified as crucial to understanding the regional patterns of disease transmission observed in data \cite{Castillo-Chavez2016-uh}. In the case of natural metapopulations, previous analyses found that dispersal can either facilitate or dilute the transmission of the disease \cite{Huang2015-or}. Facilitation of transmission can occur due to the presence of superspreader patches with high reproduction numbers \cite{Lieberthal2021-va}, while dilution of transmission can occur when patches with low reproduction numbers act as refugia where the population can persist \cite{Heard2015-jl}.

Another factor that can affect the risk of extinction due to the disease of host populations is the stage structure of this population. Pathogens may have different transmission rates or mortality effects on different life stages of a species \cite{Dwyer1991-xj}. Moreover, this can cause a shift in the stage structure and reduce the population resiliency to disturbance \cite{Hite2016-uu, Hite2023-dc}. This loss of resiliency is particularly problematic when mortality is higher in juvenile, nonreproductive stages. High mortality of juveniles reduces the number of reproductive individuals and causes a population decline even when fertility rates are high \cite{Briggs2005-qr}.

A case example of conservation interest where the risk of population extinction, spatial variation, and life history play a role in population decline is the sunflower sea star, \textit{Pycnopodia helianthoides} \cite{Hamilton2021-tn}. This species has experienced a substantial population decline in the Northeastern Pacific driven by increased water temperatures and the presence of SSWD \cite{Harvell2019-mg}. This decline has led the local population to a functional extinction in some parts of the coastline. \textit{Pycnopodia} is one of the main predators of the barren-forming purple urchin \textit{Strongylocentrotus purpuratus}. This decline has thus contributed to a substantial loss of coverage of the highly productive kelp forest in places where \textit{Pycnopodia} is the main predator due to an increased grazing pressure from purple sea urchins \cite{Rogers-Bennett2019-lq}. Furthermore, this increases the interest in recovering \textit{Pycnopodia} populations to recover kelp forest coverage \cite{Galloway2023-eu} and promote its long-term resistance to disturbance events \cite{ArroyoEsquivel-qr}.

In the northern extent of \textit{Pycnopodia}’s geographic range, the effect of SSWD has been patchy \cite{Hamilton2021-tn}. Remnant populations of \textit{Pycnopodia} have been found in nearshore relatively shallow habitats within Fjords in British Columbia following regional exposure to SSWD \cite{Hamilton2021-tn, lowry_endangered_2023}. In contrast, adjacent habitats considerably declined \cite{Hamilton2021-tn, lowry_endangered_2023}. Importantly, populations within Fjords have maintained a mean individual size above the reproductive minimum. In contrast, since the SSWD outbreak, the mean size in adjacent habitats has not reached the reproductive minimum \cite{lowry_endangered_2023}. In the years following the SSWD outbreak, there has been consistent recruitment of juvenile \textit{Pycnopodia}, likely sourced from the populations within the Fjords \cite{lowry_endangered_2023}.

This paper presents a mathematical analysis of a two-patch, two-life stages SI model with unidirectional dispersal and no recovery from infection. Two-patch models are classical approaches to analyzing spatial dynamics of metapopulations mathematically. With the increasing interest in understanding spatial dynamics of disease spread, many two-patch epidemiological models have appeared in recent literature \cite{Kang2012,Tewa12, Arino16, Calvo20,  Lee20, DongXue20, SahaSamanta22}. In the next section, we will introduce our model, which will then be followed by a mathematical and numerical analysis of its equilibria in Section 3. Finally, we will discuss the ecological implications of our analysis in Section 4.

\section{Model and methods}

We propose a two-patch SI model \cite{Kang2012} where a natural population has a density $N_i$ at patch $i=1,2$, where patch 1 represents a source patch and patch 2 is a sink patch. On each patch, healthy adult individuals $A_{S,i}$ produce healthy offspring $J_{S,i}$ following a logistic growth functional form. A proportion $p$ of juveniles stay in the patch, while the remaining $1-p$ escapes the patch, either from the source patch into the sink patch (Patch 1) or from the sink patch outside the system (Patch 2). In addition, healthy juveniles mature and become reproductive adults at a rate $\alpha$.

Healthy individuals can be infected following direct contact with an infected juvenile $J_{I,i}$ or adult $A_{I,i}$ with an age-independent, patch-specific transmission rate $\beta_i$. To model the effects of disease mortality, we separate mortality rates of healthy individuals $\mu_S$ from infected individuals $\mu_I$, where the common assumption of $\mu_I>\mu_S$ is made. Finally, we model the slow or rare recovery events by assuming that no transition from infected to susceptible individuals is possible.

In equation form, our system is written as

\begin{align}\label{eq:baseModel}
& \text{Patch 1} \begin{cases}
    \dot{J_{S,1}}=&pr\left(1- \frac{N_1}{k}\right)A_{S,1}-\beta_1J_{S,1}\frac{(J_{I,1}+A_{I,1})}{N_1}-\alpha J_{S,1}-\mu_SJ_{S,1}, \nonumber \\
    \dot{J_{I,1}}=&\beta_1J_{S,1}\frac{(J_{I,1}+A_{I,1})}{N_1}-\mu_IJ_{I,1}, \\
    \dot{A_{S,1}}=&\alpha J_{S,1}-\beta_1A_{S,1}\frac{(J_{I,1}+A_{I,1})}{N_1}-\mu_SA_{S,1},\\
    \dot{A_{I,1}}=&\beta_1A_{S,1}\frac{(J_{I,1}+A_{I,1})}{N_1}-\mu_IA_{I,1},
\end{cases} \\\\
& \text{Patch 2} \begin{cases}
        \dot{J_{S,2}}=&(1-p)r\left(1- \frac{N_2}{k}\right)A_{S,1}+pr\left(1- \frac{N_2}{k}\right)A_{S,2}-\beta_2J_{S,2}\frac{(J_{I,2}+A_{I,2})}{N_2}-\alpha J_{S,2}-\mu_SJ_{S,2},\nonumber\\
        \dot{J_{I,2}}=&\beta_2J_{S,2}\frac{(J_{I,2}+A_{I,2})}{N_2}-\mu_IJ_{I,2},\\
        \dot{A_{S,2}}=&\alpha J_{S,2}-\beta_2A_{S,2}\frac{(J_{I,2}+A_{I,2})}{N_2}-\mu_SA_{S,2},\\
        \dot{A_{I,2}}=&\beta_2A_{S,2}\frac{(J_{I,2}+A_{I,2})}{N_2}-\mu_IA_{I,2},
    \end{cases}
\end{align}

\noindent where $N_1 := J_{S,1}+ A_{S,1} + J_{I,1} + A_{I,1}$ and $N_2 := J_{S,2}+ A_{S,2} + J_{I,2} + A_{I,2}$. Consequently, (\ref{eq:baseModel}) gives us that

\begin{equation}\label{eq:Nmodel}
    \begin{split}
        \dot{N_1}=&pr\left(1- \frac{N_1}{k}\right)A_{S,1}-\mu_S(J_{S,1}+A_{S,1})-\mu_I(J_{I,1}+A_{I,1})\\
        \dot{N_2}=&(1-p)r\left(1- \frac{N_2}{k}\right)A_{S,1}+rA_{S,2}-\mu_S(J_{S,2}+A_{S,2})-\mu_I(J_{I,2}+A_{I,2})\\
    \end{split}
\end{equation}

A transfer diagram for this model is given in Figure 1.

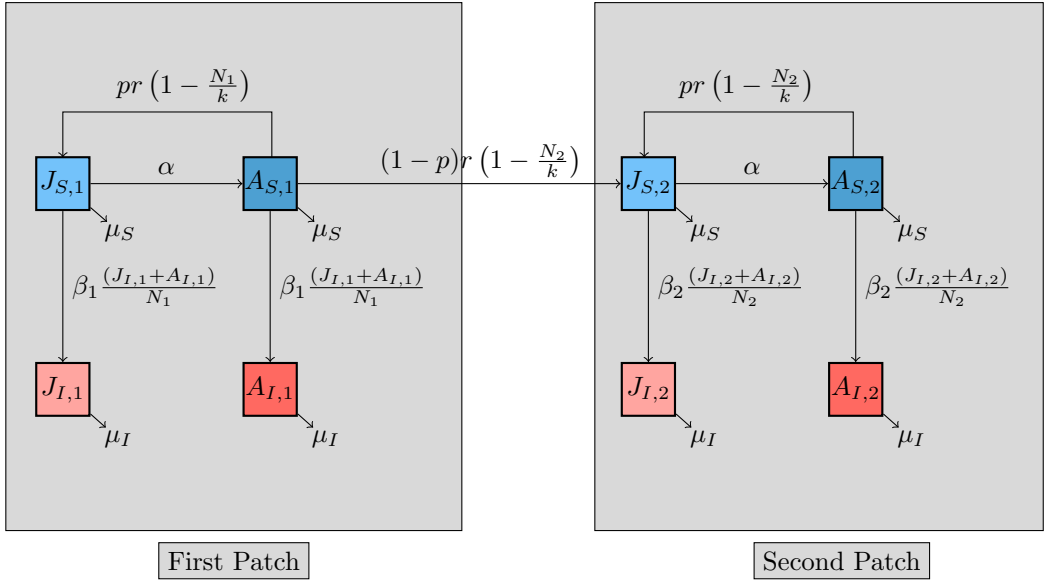
\begin{figure}[H]
    \centering
    \begin{tikzpicture}[%
            node distance = 20mm,
            state/.style = {%
                rectangle,
                        draw = black,
                   inner sep = 0pt,
                minimum size = 7mm,
                thick,
            },
            beta/.style = {%
            node distance = 2mm,
                inner sep = 1pt,
            },
            auto,
        ]

        \draw[draw=black, fill=gray!30] (-0.75,0.4) rectangle ++(6,7);
        \draw[draw=black, fill=gray!30] (7,0.4) rectangle ++(5.9,7);
        
        \node[draw, fill=gray!30] at (2.25, 0) (a) {First Patch};
        \node[draw, fill=gray!30] at (10.25, 0) (a) {Second Patch};

        \node (JS1) [state, fill=mayablue] at (0,5) {$J_{S,1}$};
        \node (JI1) [below=of JS1,state, fill=pastelred!60] {$J_{I,1}$};
        \node (AS1) [right=of JS1,state,fill=oceanboatblue!70] {$A_{S,1}$};
        \node (AI1) [below=of AS1,state,fill=pastelred] {$A_{I,1}$};

        \draw [->] (JS1) to node {$\alpha$} (AS1);
        \draw [->] (JS1) to node {$\beta_1\frac{(J_{I,1}+A_{I,1})}{N_1}$} (JI1);
        \draw [->] (AS1) to node {$\beta_1\frac{(J_{I,1}+A_{I,1})}{N_1}$} (AI1);

        \node (JS1mu) [beta,below right=of JS1] {$\mu_S$};
        \draw [->] (JS1) -- (JS1mu);

        \node (AS1mu) [beta,below right=of AS1] {$\mu_S$};
        \draw [->] (AS1) -- (AS1mu);

        \node (JI1mu) [beta,below right=of JI1] {$\mu_I$};
        \draw [->] (JI1) -- (JI1mu);

        \node (AI1mu) [beta,below right=of AI1] {$\mu_I$};
        \draw [->] (AI1) -- (AI1mu);

        \draw[->] (2.75,5.35) -- (2.75,5.95) -- (0,5.95) -- (0,5.35);
        \node[] at (1.6,6.3) {$pr\left(1- \frac{N_1}{k}\right)$};

        \node (interm) [right=of AS1] {};
        
        \node (JS2) [right=of interm,state, fill=mayablue] {$J_{S,2}$};
        \node (JI2) [below=of JS2,state, fill=pastelred!60] {$J_{I,2}$};
        \node (AS2) [right=of JS2,state,fill=oceanboatblue!70] {$A_{S,2}$};
        \node (AI2) [below=of AS2,state,fill=pastelred] {$A_{I,2}$};

        \draw [->] (JS2) to node {$\alpha$} (AS2);
        \draw [->] (JS2) to node {$\beta_2\frac{(J_{I,2}+A_{I,2})}{N_2}$} (JI2);
        \draw [->] (AS2) to node {$\beta_2\frac{(J_{I,2}+A_{I,2})}{N_2}$} (AI2);

        \node (JS2mu) [beta,below right=of JS2] {$\mu_S$};
        \draw [->] (JS2) -- (JS2mu);

        \node (AS2mu) [beta,below right=of AS2] {$\mu_S$};
        \draw [->] (AS2) -- (AS2mu);

        \node (JI2mu) [beta,below right=of JI2] {$\mu_I$};
        \draw [->] (JI2) -- (JI2mu);

        \node (AI2mu) [beta,below right=of AI2] {$\mu_I$};
        \draw [->] (AI2) -- (AI2mu);

        \draw [->] (AS1) to node {} (JS2);

        \draw[->] (10.4,5.35) -- (10.4,5.95) -- (7.65,5.95) -- (7.65,5.35);
        \node[] at (9,6.3) {$pr\left(1- \frac{N_2}{k}\right)$};

        \node[] at (5.5,5.3) {$(1-p)r\left(1-\frac{N_2}{k}\right)$};

    \end{tikzpicture}
    \caption{Transfer diagram for model (\ref{eq:baseModel}).}
    \label{fig:model_transfer_diagram}
\end{figure}

\begin{remark}
In the following section, we explore the analytical properties of this model, particularly concerning the existence and stability criteria of equilibrium points. We note that the general Jacobian matrix of this model is a block triangular matrix of the form
\begin{align*}
    \mathcal{J} = \begin{pmatrix}
\mathcal{J}_1 & \mathbf{0} \\ \mathcal{G} & \mathcal{J}_2
\end{pmatrix},
\end{align*}
where $\mathcal{J}_1, \mathcal{J}_2$ are the Jacobian matrices for the first and second patches correspondingly. In light of this, the stability study will be performed separately for each population patch.
\end{remark}


\section{Mathematical Analysis}


\subsection{First Patch Analysis}

In this section, we investigate the stability and point values for the possible equilibria of our primary model in the first patch. Note that this patch is entirely independent of the second patch; hence, starting our analytical study with this patch alone makes sense. There are three main types of equilibrium points possible for this patch: the \textbf{disease-free} equilibrium, in which the population of infected, $I_1 = J_{I,1}+A_{I,1}$ equals zero, the \textbf{endemic} equilibrium, in which $I_1$ attains a nonzero value, and the \textbf{extinction} equilibrium, in which all values, $J_{S,1},A_{S,1},J_{I,1}$ and $A_{I,1}$ converge to zero (the population becomes extinct). We investigate the possible parameters for each scenario that control the point's stability.

\subsubsection{Disease-free equilibrium for the first patch}

After some calculations, if $\mathcal{J}_1$ is the Jacobian matrix for the first patch, then we have that $ \det(\mathcal{J}_1 - \lambda I)$ is given by 
\begin{small}
    \begin{align}\label{eq:determinant}
    \det \begin{bmatrix}
        - \frac{prA_{S,1}}{k} - \beta_1 \delta_1 + \beta_1J_{S,1} \eta_1 - (\alpha+\mu_S) -\lambda & - \frac{prA_{S,1}}{k} - \beta_1 J_{S,1}\vartheta_1 &  - \frac{2prA_{S,1}}{k} + \beta_1 J_{S,1}\eta_1 & - \frac{prA_{S,1}}{k} - \beta_1 J_{S,1}\vartheta_1 \\
        \beta_1 \delta_1 - \beta_1 J_{S,1} \eta_1 & \beta_1 J_{S,1} \vartheta_1 - \mu_I -\lambda & -\beta_1 J_{S,1}\eta_1 & \beta_1 J_{S,1} \vartheta_1 \\
        \alpha + \beta_1 A_{S,1} \eta_1 & - \beta_1 A_{S,1} \vartheta_1 & - \beta_1 \delta_1 + \beta_1 A_{S,1}\eta_1 - \mu_S- \lambda & - \beta_1A_{S,1}\vartheta_1 \\
        -\beta_1 A_{S,1} \eta_1 & \beta_1 A_{S,1}\vartheta_1 & \beta_1 \delta_1 - \beta_1 A_{S,1} \eta_1 & \beta_1 A_{S,1} \vartheta_1 - \mu_I -\lambda
    \end{bmatrix}
\end{align}
\end{small}
where 
\begin{align*}
    \delta_1: = \frac{J_{I,1}+A_{I,1}}{N_1}, \quad \eta_1 := \frac{J_{I,1}+A_{I,1}}{N_1^2}, \quad \vartheta_1: = \frac{J_{S,1}+A_{S,1}}{N_1^2}.
\end{align*}
When we find ourselves in the disease-free equilibrium for this patch, we have $J_{I,1}=A_{I,1}=0$, implying that $\delta_1=\eta_1=0$ and $\vartheta_1 = \frac{1}{N_1}$. Substituting these values in the determinant (\ref{eq:determinant}) we must compute

\begin{small}
\begin{equation}\label{eq:first_patch_dfe_jacobian}
\det \begin{bmatrix}
        - \frac{prA_{S,1}}{k} - (\alpha+\mu_S) -\lambda & - \frac{prA_{S,1}}{k} - \beta_1 \frac{J_{S,1}}{N_1} &  - \frac{2prA_{S,1}}{k} & - \frac{prA_{S,1}}{k} - \beta_1 \frac{J_{S,1}}{N_1} \\
        0 & \beta_1 \frac{J_{S,1}}{N_1} - \mu_I -\lambda & 0 & \beta_1 \frac{J_{S,1}}{N_1} \\
        \alpha & - \beta_1 \frac{A_{S,1}}{N_1} & - \mu_S- \lambda & - \beta_1\frac{A_{S,1}}{N_1} \\
        0 & \beta_1 \frac{A_{S,1}}{N_1} & 0 & \beta_1 \frac{A_{S,1}}{N_1} - \mu_I -\lambda
    \end{bmatrix}.
\end{equation} 
\end{small}
From section \ref{sec:first_patch_dfe_jacobian} in the appendix, we obtain that this determinant equals
\begin{align*}
    \left[- \left( - \frac{prA_{S,1}}{k} - (\alpha+\mu_S) -\lambda \right) \cdot (\mu_S + \lambda) + \alpha \cdot \frac{2prA_{S,1}}{k} \right] \left[ - \frac{\beta_1}{N_1}(\mu_I+ \lambda)(A_{S,1}+J_{S,1}) + (\mu_I+ \lambda)^2\right],
\end{align*}
which is a product of two quadratic polynomials. We simplify both of them. For the second polynomial, note that when the system attains the disease-free-equilibrium we have that $N_1 = J_{S,1}+A_{S,1}$ so the second polynomial becomes

\begin{align*}
    &- \frac{\beta_1}{N_1}(\mu_I+ \lambda)(A_{S,1}+J_{S,1}) + (\mu_I+ \lambda)^2 = - \beta_1(\mu_I+ \lambda) + (\mu_I+ \lambda)^2 = \\
    &(\mu_I+ \lambda)(-\beta_1 + \mu_I + \lambda) = (\mu_I+ \lambda)( \lambda - (\beta_1 - \mu_I)).
\end{align*}

The reader can verify that the first polynomial is of the form $\lambda^2 + B_1\lambda + C_1$ where $B_1>0$ and $C_1>0$, so both of it's roots are negative \footnote{Indeed, in equation $ax^2+bx+c=0,$ where $a,b,c>0$, then one possible root has numerator $-b-\sqrt{b^2-4ac}$ which is negative, and the other has numerator $-b+\sqrt{b^2-4ac}$ which is also negative: $-b+ \sqrt{b^2-4ac}<0 \Leftrightarrow \sqrt{b^2-4ac} < b \Leftrightarrow b^2 - 4ac < b^2 \Leftrightarrow 0 <4ac.$}. In conclusion, the characteristic polynomial of the Jacobian matrix at the disease-free equilibrium is of the form
$$
(\lambda^2 + B_1\lambda + C_1)(\mu_I+ \lambda)( \lambda - (\beta_1 - \mu_I)),
$$
with the following eigenvalues: $\lambda_1, \lambda_2$ (the two negative roots of  $\lambda^2 + B_1\lambda + C_1$), - $\mu_I$ and $\beta_1 - \mu_I$. Consequently, the disease-free equilibrium in this patch is stable when $\beta_1 < \mu_I$. This leads us to define the \textbf{basic reproductive number for the first patch}, given by \begin{equation}\label{eq:R01}R_{0,1} : = \frac{\beta_1}{\mu_I}.\end{equation} We then have the following.

\begin{proposition}
    When $R_{0,1}<1$, the disease-free equilibrium in the first patch is stable.
\end{proposition}

\begin{remark}
    The disease-free equilibrium can be computed explicitly as well, and a simple calculation shows that it is given by
    \begin{align}\label{eq:diseasefreeeqP1}
       A_{S,1}^{*,0} = \frac{k \left[ 1 - \frac{\mu_S}{\alpha p r} (\mu_S+\alpha) \right]}{1+ \frac{\mu_S}{\alpha}} , \quad  J_{S,1}^{*,0} = \frac{\mu_S}{\alpha}A_{S,1}^{*,0} , \quad  
       A_{I,1}^{*,0} = 0, \quad  
       J_{I,1}^{*,0} = 0,
    \end{align}
    if $A_{S,1}^{*,0} \geq 0.$
\end{remark}

\subsubsection{Endemic equilibrium for the first patch} Consider the variable $I_1:= J_{I,1}+A_{I,1}$, the total infected population in the first patch of the model. We find the existence of equilibria where $I_1>0$. The appendix's calculations in section \ref{sec:first_patch_ee_calculation} give us the following proposition regarding this endemic equilibria.

\begin{proposition}\label{prop:endemic_equilibrium_patch_1}
    If the first patch of the model (\ref{eq:baseModel}) attains an endemic equilibrium, then it is given by 
    \begin{align}\label{eq:endemic_equilibrium_values1}
    J_{S,1}^* = \frac{1}{(R_{0,1}-1)(\lambda_1+1)}I_1^*, \quad 
    A_{S,1}^* = \frac{\lambda_1}{(R_{0,1}-1)(\lambda_1+1)}I_1^*, \\
    J_{I,1}^* = \frac{1}{\lambda_1+1}I_1^*, \quad
    A_{I,1}^* = \frac{\lambda_1}{\lambda_1+1}I_1^*,
    \end{align}
    where
    \begin{equation}\label{eq:subModel8_2_A}
    I_1^* = \frac{k\left(R_{0,1}-1\right)}{r R_p R_{\alpha} R_{0,1}}\biggl(rR_pR_{\alpha} - (R_{0,1} + R_s + R_{\alpha} -1)(R_{0,1} + R_s -1)\biggr),
    \end{equation}
    and
    \begin{equation}\label{eq:R_definitions_A}
    R_p := \frac{p}{\mu_I}, \quad R_{\alpha} := \frac{\alpha}{\mu_I}, \quad R_s := \frac{\mu_S}{\mu_I}, \quad 
    \lambda_1 := \frac{R_{\alpha}}{R_{0,1} + R_s - 1}.
    \end{equation}
    Provided that $I_1^*>0$ and that $R_{0,1}>1$ \footnote{When $R_{0,1}<1$ we saw, previously, that the disease-free equilibrium is stable.}.
\end{proposition}

\begin{remark}\label{rem:roots_patch_1}
    Let us carefully analyze Equation (\ref{eq:subModel8_2_A}). The reader will note that, looking at $R_{0,1}$ as a variable, there are three roots for $I_1^*=0$, given by:
    \begin{align*}
        & R_{0,1} = 1, \\
        & R_{0,1} = 1 + \frac{R_{\alpha}}{2}\left( -1 - \sqrt{1+4r \frac{R_p}{R_\alpha}}\right) - R_s = 1 - \left[\frac{R_{\alpha}}{2}\left(1 + \sqrt{1+4r \frac{R_p}{R_\alpha}}\right) + R_s\right], \\
        & R_{0,1} = 1 + \frac{R_{\alpha}}{2}\left( -1 + \sqrt{1+4r \frac{R_p}{R_\alpha}}\right) - R_s.
    \end{align*}
    The second root is always less than $1$. However, the third root can be greater or less than $1$. Both cases yield very different situations, as shown in our numerical simulations below.
\end{remark}

\subsubsection{Extinction equilibrium for the first patch} Another type of equilibrium that might be present in this model is the extinction equilibrium. It is simple to verify that the determinant at (\ref{eq:determinant}), when $A_{S,1}=J_{S,1}=A_{I,1}=J_{I,1}=0$ equals $(\mu_I+ \lambda)^2(\mu_S+\lambda)(\lambda + \alpha+ \mu_S)$, which has all negative roots. Moreover, this implies that the extinction equilibrium is stable. This equilibrium will be attained when $R_{0,1}>1$ and $I_1^*$ in equation (\ref{eq:subModel8_2_A}) is negative. Again, this will be discussed more in-depth in our numerical results section.


\subsection{Second Patch Analysis} We now continue with a similar analysis of the equilibria possibilities for the second patch. Note that this patch is dependent on the first due to the presence of the term $A_{S,1}$ in the first equation of the second patch. Consequently, results for this patch will also depend on the equilibrium point attained by the first patch.

\subsubsection{Disease-free equilibrium for the second patch}

The Jacobian matrix of the second patch, $\mathcal{J}_2$, is analogous to $\mathcal{J}_1$, with the exception that each entry of the first row has the addition of the extra term $-\frac{(1-p)r}{k}A_{S,1}$. Calculations made in section \ref{sec:second_patch_dfe_jacobian} of the appendix assure us that the determinant $\det(\mathcal{J}_2 - \lambda I)$ is again of the form 

$$
(\lambda^2 + B_2\lambda + C_2)(\mu_I+ \lambda)( \lambda - (\beta_2 - \mu_I)),
$$
where $B_2, C_2>0$. This yields a similar situation as in the first patch, where the disease-free equilibrium in the second patch is stable when $\beta_2 < \mu_I$. Again this obliges us to define the \textbf{basic reproductive number for the second patch}, given by \begin{equation}\label{eq:R02}
    R_{0,2}:= \frac{\beta_2}{\mu_I}.
\end{equation} We have thus proved the following.

\begin{proposition}
    When $R_{0,2}<1$, the disease-free equilibrium in the second patch is stable.
\end{proposition}

\subsubsection{Endemic equilibria for the second patch}

The endemic equilibria in the second patch are those equilibria points in which $I_2 := J_{I,2}+A_{I,2}$ is positive. The only difference with the corresponding analysis for the first patch consists of an extra term that depends on $A_{S,1}$ and $N_2$ for the first equation in this patch. We recall that this equation is given at the equilibrium point by
\begin{align}\label{eq:endemic_eq_2_raw_eq}
    0 = (1-p)r\left(1- \frac{N_2}{k}\right)A_{S,1}+pr\left(1- \frac{N_2}{k}\right)A_{S,2}-\beta_2J_{S,2}\frac{(J_{I,2}+A_{I,2})}{N_2}-\alpha J_{S,2}-\mu_SJ_{S,2}.
\end{align}
Using $I_2 = J_{I,2}+A_{I,2}$ we can repeat the same analogous from section \ref{sec:first_patch_ee_calculation}, applied to the second patch, to obtain the relations
\begin{align}\label{eq:endemic_equilibrium_values2}
    & J_{S,2}^* = \frac{1}{(R_{0,2}-1)(\lambda_2+1)}I_2^*, \quad 
    A_{S,2}^* = \frac{\lambda_2}{(R_{0,2}-1)(\lambda_2+1)}I_2^*, \nonumber \\
    & J_{I,2}^* = \frac{1}{\lambda_2+1}I_2^*, \quad A_{I,2}^* = \frac{\lambda_2}{\lambda_2+1}I_2^*,  \quad N_2^* = \frac{R_{0,2}}{R_{0,2}-1}I_2^* 
\end{align}
and $N_2^* = \frac{R_{0,2}}{R_{0,2}-1}I_2^*$, where
\begin{align}\label{eq:lambda_2_definition}
    \lambda_2 := \frac{R_{\alpha}}{R_{0,2} + R_s - 1}.
\end{align}
We can modify Equation (\ref{eq:endemic_eq_2_raw_eq}) to obtain the quadratic equation defining the equilibrium $I_2^*$. We state this in the following proposition.

\begin{proposition}\label{prop:endemic_equilibrium_patch_2}
    When the first patch attains an equilibrium ($J_{S,1}^*$, $A_{S,1}^*$, $J_{I,1}^*$, $A_{I,1}^*$), and the second patch has endemic equilibria, they are given by the relations in (\ref{eq:endemic_equilibrium_values2}), where $I_2^*$ is a solution of the quadratic equation
    \begin{align}\label{eq:quadeqP2}
    0 &= \frac{pr \lambda_2 R_{0,2}}{k(\lambda_2+1)(R_{0,2}-1)^2} (I_2^*)^2  \\
    & + \left[ 
    \frac{1}{(\lambda_2+1)(R_{0,2}-1)}\left(-pr\lambda_2 + \frac{\beta_2 (R_{0,2}-1)}{R_{0,2}} + \alpha + \mu_S \right) + \frac{(1-p)rR_{0,2}A_{S,1}^*}{k(R_{0,2}-1)}
    \right]I_2^* \nonumber \\
    & - (1-p)rA_{S,1}^* \nonumber.
    \end{align}
    and $\lambda_2$ is given by Equation \ref{eq:lambda_2_definition}.
\end{proposition}

\section{Numerical Analysis}


\begin{example}[First Patch Analysis]

Remark \ref{rem:roots_patch_1} indicates that the analysis of the equilibrium of the first patch depends highly on the values of the roots of the cubic polynomial that defines $I_1^*$ in terms of the basic reproductive number for the first patch, $R_{0,1}$. The following graphs show disease-free and endemic equilibria at the first patch when the third root of this polynomial is greater than $1$. We highlight the type of equilibrium attained at each region.

\begin{figure}[H]
\centering
\includegraphics[scale=0.8]{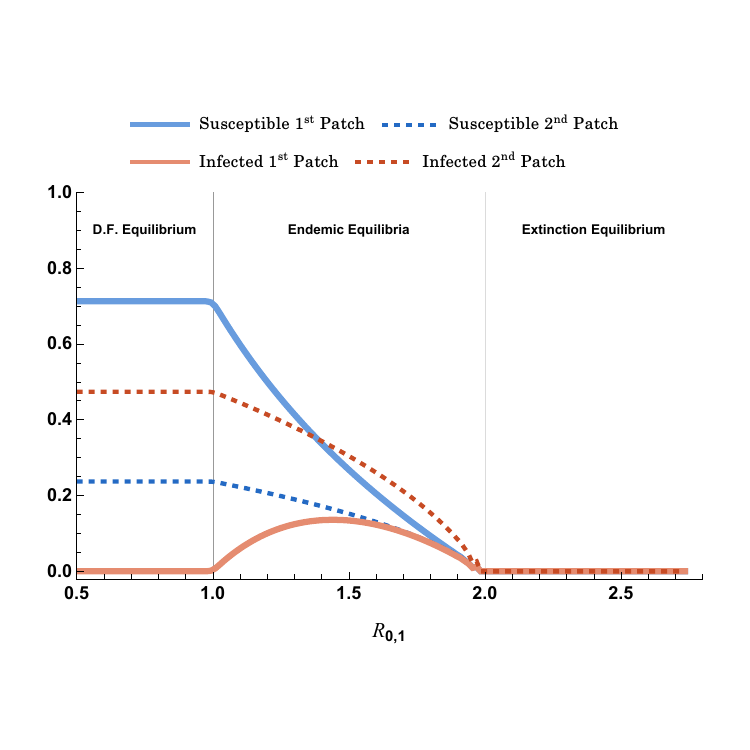}
\caption{Final infected (red) and susceptible (blue) proportions attained in both patches, varying values of $R_{0,1}$. Keeping the value of $R_{0,2}$ constant, with $R_{0,2}>1$. Note that when $R_{0,1}<1$, the disease-free equilibrium for the first patch is stable. When $R_{0,1}>1$, this equilibrium is not stable; this leads to an endemic equilibrium, given by the theoretical value of $I_1^*$ in proposition (\ref{prop:endemic_equilibrium_patch_1}). However, for larger values of $R_{0,1}$, the value of $I_1^*$ becomes negative, so the system converges to the \textbf{extinction} equilibrium.}
\label{fig:inf_eq_vs_beta_for_mu_I}
\end{figure}

In the scenario of Figure (\ref{fig:inf_eq_vs_beta_for_mu_I}), the third root (for $I_1^*=0$) of the polynomial in Equation (\ref{eq:subModel8_2_A}) is greater than $1$, giving us an interval $1<R_{0,1}<1+\epsilon$ in which $I_1^*>0$, which yields the region for the endemic equilibrium. When the third root is less than $1$, no such interval exists, and the system only has two possibilities: if $R_0<1$, it attains a disease-free equilibrium, and if $R_0>1$, it attains the extinction equilibrium. This case is rather extreme and can be achieved, for example, with very low values of the parameter $\alpha$.

\end{example}


\begin{example}[Second Patch Analysis]

While the first patch is independent and only affected by its reproductive number $R_{0,1}$, the second patch is affected by the behavior of both patches as an aggregate. In this situation, the equilibria attained by the second patch are influenced by the value of $R_{0,1}$ and $R_{0,2}$. Figure (\ref{fig:r1-r2-effect-infected-patch-2}) shows this relationship. Each point represents the average equilibrium point attained by the system using the corresponding values of $(R_{0,1}, R_{0,2})$ for several simulations with different initial conditions. With disease parameters $r = 1, k = 1, \alpha = 0.6, \mu_S = 0.1, \mu_I = 0.1$ and $
p = 0.5$. As shown before, when $R_{0,2}<1$, the disease-free equilibrium is stable in the second patch. When $R_{0,2}>1$ the equilibrium depends on both $R_{0,1}$ and $R_{0,2}$. The region on the bottom right represents an extinction scenario for larger values of both $R_{0,1}$ and $R_{0,2}$. A range of $R_{0,2}$ values yields endemic equilibria in the second patch, regardless of the $R_{0,1}$ value.

\begin{figure}[H]
    \centering
    \includegraphics[scale=0.6]{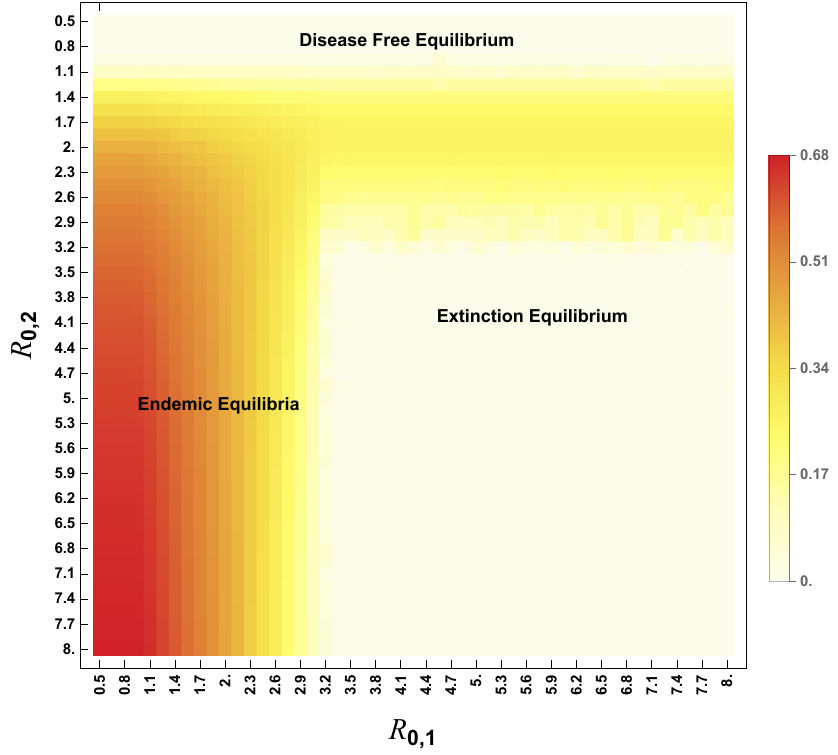}
    \caption{Final infected proportion in second patch for each $(R_{0,1}, R_{0,2})$ combination and an illustration of the three different types of equilibria regions for each case.}
    \label{fig:r1-r2-effect-infected-patch-2}
\end{figure}

\end{example}

\begin{example}[Stability of the general endemic equilibrium]
Computing the eigenvalues of $\mathcal{J}_1$ and $\mathcal{J_2}$ at the corresponding endemic equilibria becomes an arduous process. To inspect the stability of these points, we perform simulations and present the maximum eigenvalues of these Jacobian matrices at the endemic equilibrium scenarios. The results of this experiment are presented in Figure \ref{fig:max-eigs-figures} for each patch. When $R_{0,1}<1$ or $R_{0,2}<1$, the corresponding disease-free equilibria are stable (as proved before). When $R_{0,1}$ and $I_1^*>0$ according to Equation (\ref{eq:subModel8_2_A}), then the corresponding endemic equilibrium seems to be also stable for all simulations performed. In the ultimate case, the extinction equilibrium is also stable. 

\begin{figure}[H]
    \centering
    \begin{tabular}{cc}
    (a) & \includegraphics[scale=0.5]{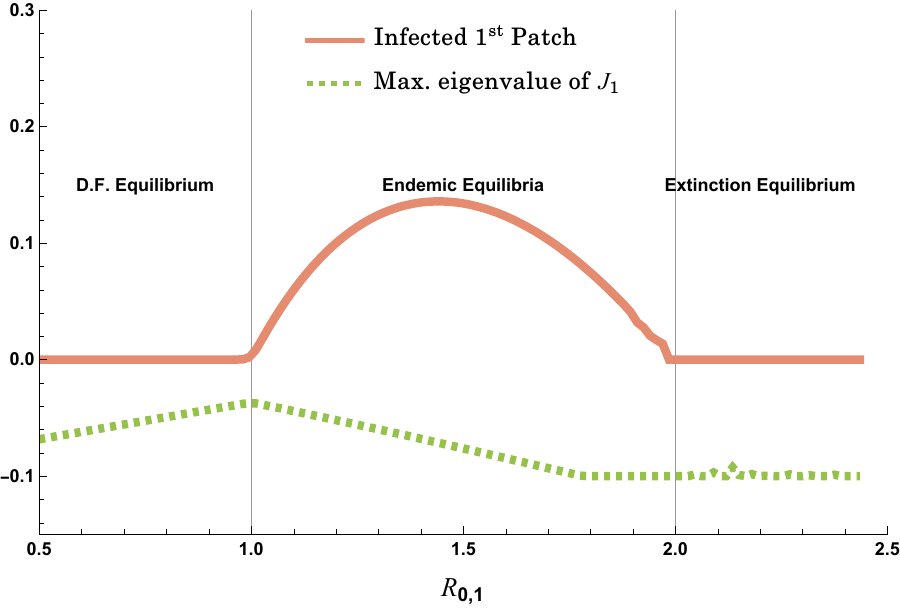} \\
    (b) & \includegraphics[scale=0.5]{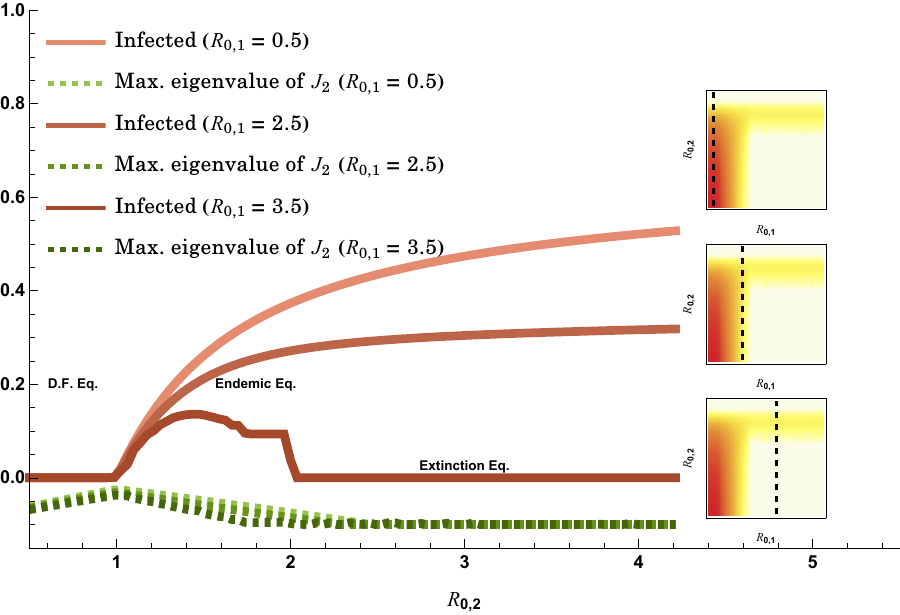}
    \end{tabular}
    \caption{(a): Maximum eigenvalue of $\mathcal{J}_1$ for each $R_{0,1}$ along with the infected population attained at the corresponding equilibrium point. We can see that all equilibria are stable in their corresponding $R_{0,1}$ regions. (b): Maximum eigenvalue of $\mathcal{J}_2$ for each $R_{0,2}$, with different curves for three values of $R_{0,1}$. Each curve corresponds to a vertical cut from Figure (\ref{fig:r1-r2-effect-infected-patch-2}).}
    \label{fig:max-eigs-figures}
\end{figure}

\end{example}

\begin{example}[Equilibria in terms of other variables]
Equation (\ref{eq:subModel8_2_A}) serves as a basis to determine the type of equilibrium that is to be achieved by the system. When $I_1^*<0$, the endemic equilibrium is not feasible, so the system will either attain a disease-free or an extinction equilibrium, depending on the values of the reproductive numbers $R_{0,1}$ and $R_{0,1}$. Equation (\ref{eq:subModel8_2_A}) depends on virtually all disease parameters, and many yield interesting bifurcation results. 

Figure (\ref{fig:ppalphagraph}) shows the effects of $p$ and the maturation rate $\alpha$ as bifurcation parameters for this system. The full impact of $\alpha, p$ and $R_{0,2}$ into the infected equilibrium of patch two is displayed in Figure (\ref{fig:ppalphagraph_matrix}). We note that these parameters determine the existence of each equilibrium type. In general, low values of $p$ and $\alpha$, or $p$ close to $1$, are cases that lead to the extinction equilibrium for the second patch.

\begin{figure}[H]
    \centering
    \begin{tabular}{cc}
    \includegraphics[width=70mm]{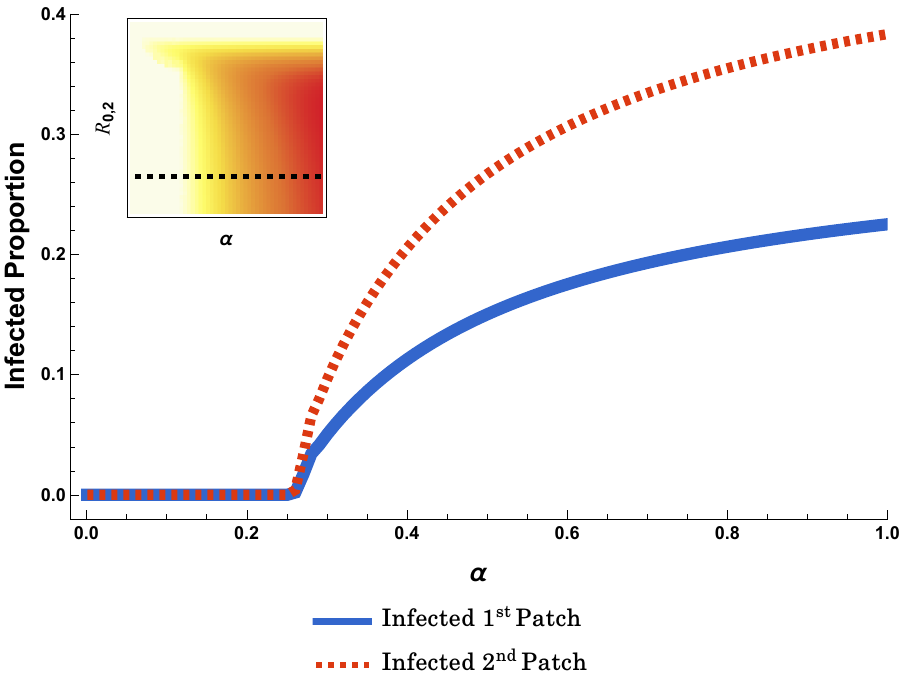} &   \includegraphics[width=70mm]{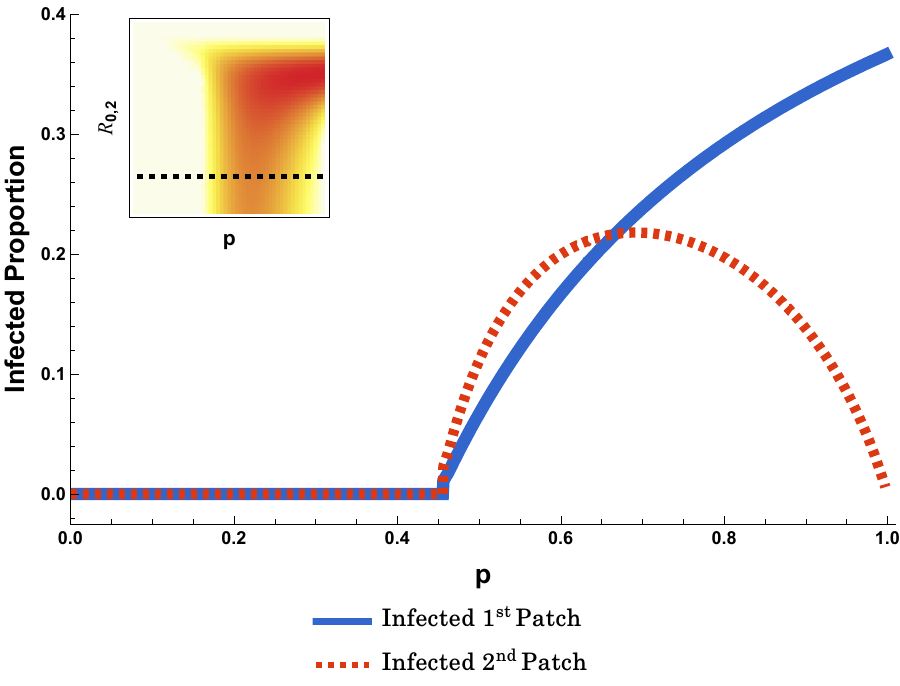} \\
    \end{tabular}
    \caption{Infected at equilibria for both patches, depending on $\alpha$ (left) and $p$ (right). Other parameters fixed: $r=1, k=1, \beta_1=0.3, \beta_2=0.6, \mu_S=0.1, \mu_I=0.1$, $\alpha=0.6$ (for right panel) and $p=0.5$ (for left panel). Each graph is shown to be a horizontal cut of the corresponding heatmaps in Figure (\ref{fig:ppalphagraph_matrix}). }
    \label{fig:ppalphagraph}
\end{figure}

\begin{figure}[H]
    \centering
    \begin{tabular}{cc}
    \includegraphics[width=70mm]{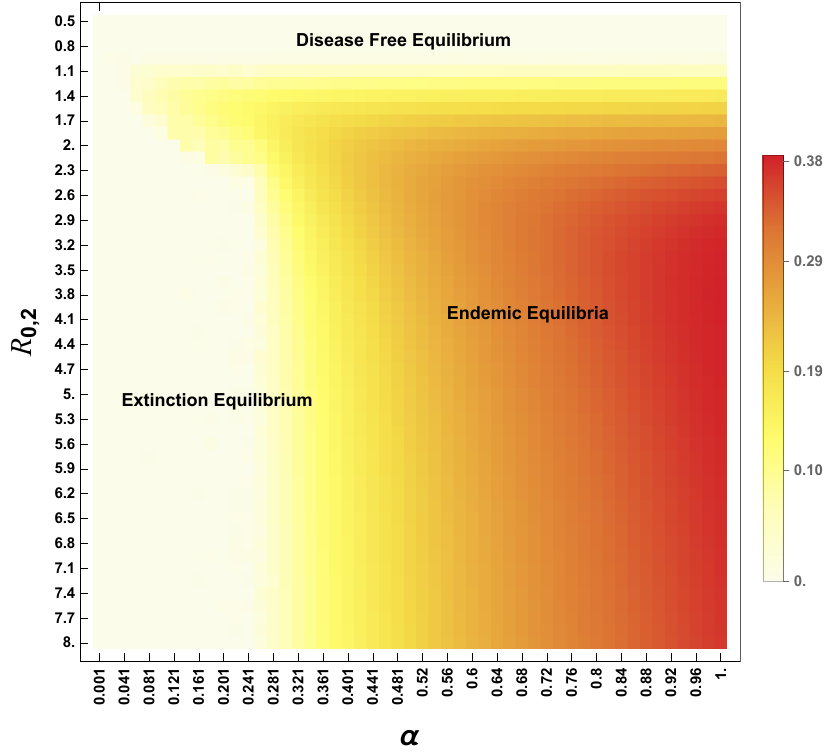} &   \includegraphics[width=70mm]{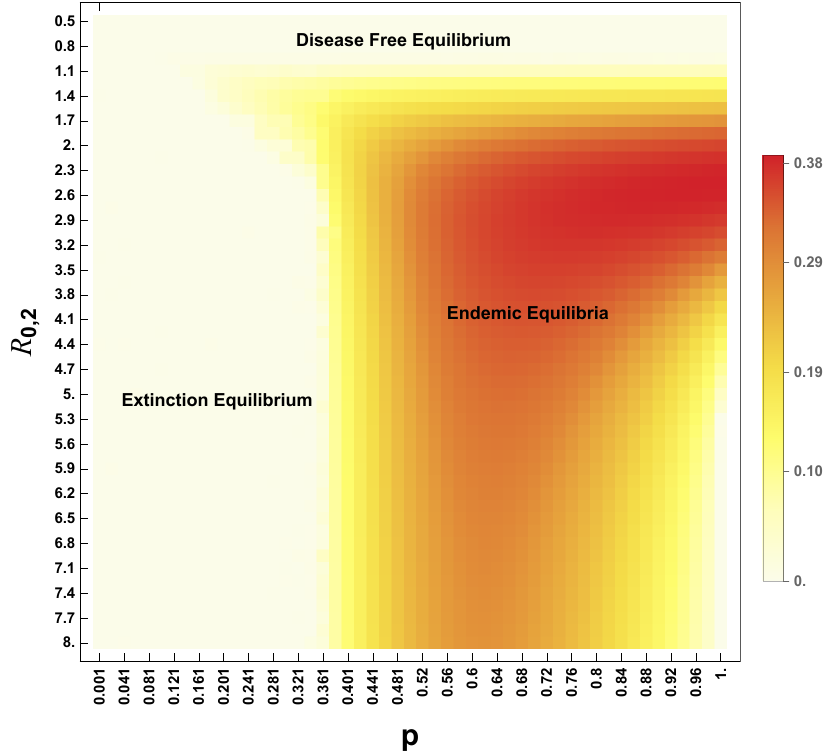} \\
    \end{tabular}
    \caption{Infected at equilibria for second patch, depending on $\alpha$ (left) and $p$ (right), varying the values of $R_{0,2}$, keeping fixed $\alpha=0.6$ (for right panel) and $p=0.5$ (for left panel).}
    \label{fig:ppalphagraph_matrix}
\end{figure}

\end{example}

\begin{example}[Population dispersion and maturation analysis] We study the juvenile to adult proportion $J_{S,2}/A_{S,2}$ for the second patch, attained at the equilibrium points. From Equation (\ref{eq:endemic_equilibrium_values2}) we obtain that
\begin{equation}\label{eq:prop_juv_adult_2_ee}
    \frac{J_{S,2}^*}{A_{S,1}^*} = \frac{1}{\lambda_2} = \frac{R_{0,2} + R_s -1}{R_{\alpha}},
\end{equation}
for the endemic equilibria case, whereas in the disease equilibrium, it is easy to compute that
\begin{equation}\label{eq:prop_juv_adult_2_dfe}
    \frac{J_{S,2}^*}{A_{S,1}^*} = \frac{\mu_S}{\alpha} = \frac{R_s}{R_{\alpha}}.
\end{equation}
We note that these equations depend inversely on the maturation rate $\alpha$ and are independent of the population dispersion $p$. However, as seen in Figure (\ref{fig:ppalphagraph_matrix}), both $p$ and $\alpha$ play a role in deciding which equilibrium is attained by the system (they are bifurcation parameters). Figure (\ref{fig:AlphaPeffect}) shows simulations where the effect of both $p$ and $\alpha$ on the ratio $\frac{J_{S,2}^*}{A_{S,1}^*}$ is displayed (which is set to the value of $0$ for the extinction equilibrium). Indeed, we can see that for endemic equilibria, the value of this ratio depends inversely on $\alpha$ and is independent of $p$.

\begin{figure}[H]
    \centering
    \includegraphics[scale=0.5]{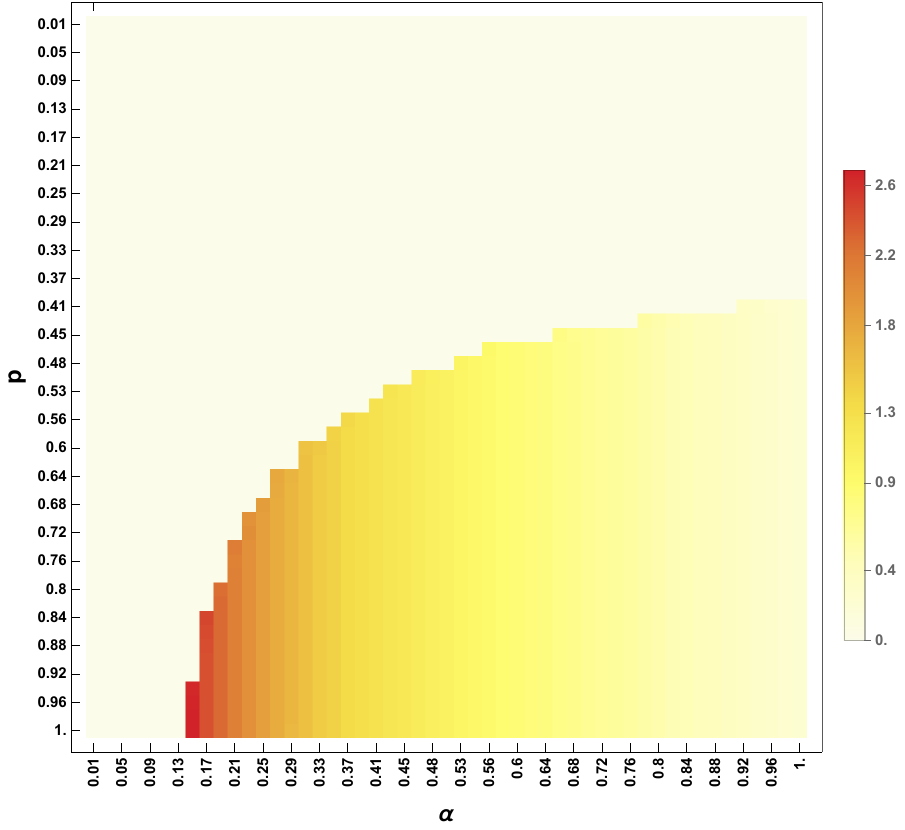}
    \caption{Effect of both $p$ and $\alpha$ on the ratio $\frac{J_{S,2}^*}{A_{S,1}^*}$ for the second patch. The ratio is set to $0$ for the extinction equilibrium (when $A_{S,1}^* = J_{S,1}^* = 0$). Using $\beta_1 = 0.3, \beta_2 = 0.4, r=k=1, \mu_S= \mu_I=0.1$. Thus having $R_{0,1} = 3$ and $R_{0,2} = 4$ (which gives the two possibilities on equilibria: endemic or extinction).}
    \label{fig:AlphaPeffect}
\end{figure}

\end{example}

\begin{example}
    Inspection of Equation (\ref{eq:quadeqP2}). The infected population at the second patch depends, as evidenced in Equation \ref{eq:quadeqP2}, on the attained number of healthy adults in the first patch. Figure \ref{fig:AS1Graph} shows the effect of $A_{S,1}^*$ over the attained value of $I_2^*$ for different values of $R_{0,2}$. The effect of $A_{S,1}^*$ becomes more important as $R_{0,2}$ increases. 
    \begin{figure}[H]\label{fig:AS1Graph}
    \centering
    \includegraphics[scale=0.6]{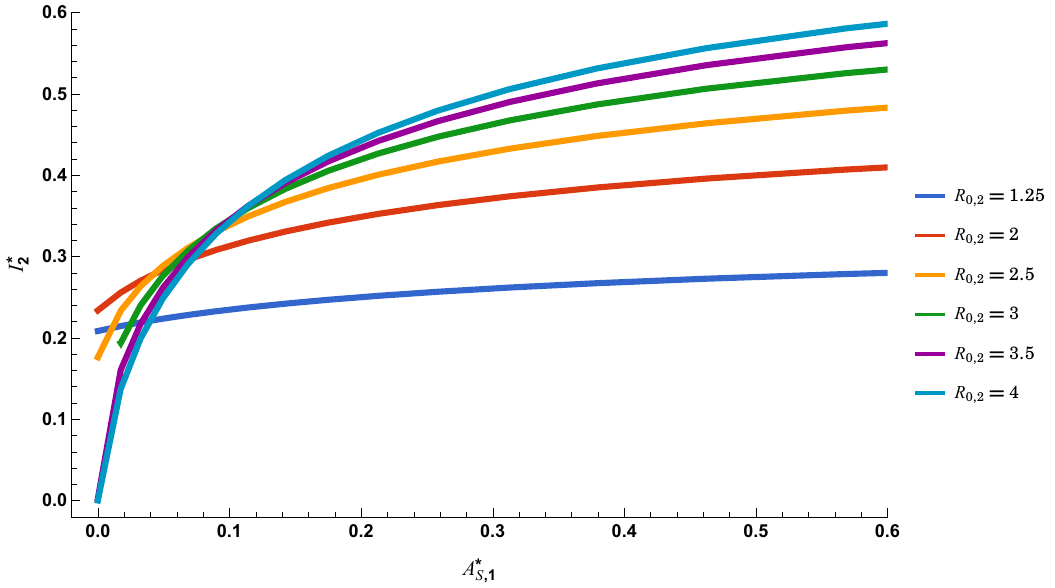}
    \caption{Infected at second patch depending on the value of $A_{S,1}^*$. This relationship is given by the quadratic Equation (\ref{eq:quadeqP2}), which depends on $A_{S,1}^*$ and $R_{0,2}$.}
    \label{fig:AS1Graph}
\end{figure}
\end{example}

\section{Discussion}
In this paper, we explored the different dynamics of a two-patch metapopulation with a disease where recovery is impossible, and dispersal occurs in a single direction from a source population to a sink population. We mathematically proved the different fates of the sink population as a function of the independent basic reproduction numbers for the source $R_{0,1}$ (Equation \ref{eq:R01}) and sink populations $R_{0,2}$ (Equation \ref{eq:R02}). We numerically explored these different fates (Figure \ref{fig:r1-r2-effect-infected-patch-2}) and found that when both basic reproduction numbers are smaller than 1, a global disease-free equilibrium occurs. As the basic reproduction numbers for the sink population $R_{0,2}$ crosses the one threshold, the sink population presents an endemic equilibrium, where the population coexists with the disease. This is the only scenario we numerically found for intermediate values of $R_{0,2}>1$. However, for higher $R_{0,2}$ values, the sink population reaches extinction if the source population is highly infectious by having a high basic reproduction number $R_{0,1}$. 

The endemic equilibrium we found in our model represents a coexistence between the host population and its disease. This coexistence can occur through several mechanisms. Our results are consistent with mechanisms such as increased resistance and tolerance to the pathogen and the presence of environmental refugia \cite{Brannelly2021-wx}. As the resistance and tolerance to the pathogen in a given patch $i$ increases, its transmission rate $\beta_i$ decreases, and thus its basic reproduction number $R_{0,i}$ decreases as well. This increases the likelihood of the sink population escaping the extinction equilibrium. A source patch with a healthy population can also act as a refugia where the sink population is sustained through recruitment inputs. This can allow the sink population to persist as the proportion of juveniles that go to the sink patch $p$ increases over a certain threshold at high basic reproduction number $R_{0,2}$ values. 

However, whether the juveniles from the source patch can produce a self-sustaining population in the sink patch or the population is only sustained by the recruitment inputs depends on how many productive adults are in the sink population at equilibrium. Looking at the stage structure of the equilibrium population in the sink patch, measured by its ratio of juveniles to adults $\frac{J_{S,2}^*}{A_{S,2}^*}$ hints at whether this population is dominated by productive adults or juveniles, potentially arriving from the source population to die (Figure \ref{fig:AlphaPeffect}). We found that when an endemic equilibrium is reached in the sink patch, the likelihood of having an ecological rescue, where the juveniles manage to become productive adults (i.e., a smaller juvenile to adult ratio) depends only on the maturation rate $\alpha$ and not on the proportion of juveniles that reach the sink patch $p$. This is consistent with other empirical and modeling efforts that have shown a higher production rate of reproductive individuals promotes eco-evolutionary rescue \cite{Muths2011-tr,Golas2021-ya}.

In the study, the infection dynamics in the second patch of a two-patch SI epidemic model are closely tied to the variable $A_{S,1}^*$, representing the number of susceptible adults in the first patch. The relationship between the infection level in the second patch and $A_{S,1}^*$ is governed by the quadratic equation (\ref{eq:quadeqP2}). This equation underscores how the infection level in the second patch is influenced not only by the internal dynamics of that patch but also significantly by the number of susceptible adults in the first patch. This inter-patch dependency is further modulated by the basic reproductive number $R_{0,2}$ of the second patch, indicating a complex interplay between the two patches. This aspect of the model underscores a critical insight into the dynamics of infectious diseases in multi-patch environments, highlighting the necessity of considering inter-patch interactions for a comprehensive understanding of disease spread and control in spatially structured populations.

With the wide geographical spread of SSWD and the lack of documented resistance to disease in \textit{Pycnopodia}, the question of whether these populations could recover is open. These modeling results suggest that reproductively viable refuge populations within a region could sustain nearby populations, assuming that host reproductive output is high enough. While the models were developed with natural populations in mind, these modeling results could inform the efficacy of different management actions suggested in the roadmap to recovery for the \textit{Pycnopodia} \cite{heady_roadmap_2022}. In particular, a captive breeding program intending to outplant lab-raised \textit{Pycnopodia} would create similar unidirectional recruitment/outplanting dynamics. With the continued low level of SSWD infection found in the surviving sea star populations across the range \cite{kay_reciprocal_2019}, outplanting will be conducted where the disease is still present in the environment.

Additionally, our model suggests increased infected mortality with strategies such as selective culling. However, this has not been found to improve infection rates in natural populations and compensate for the disease mortality \cite{LACHISH2010,Prentice2019}. With \textit{Pycnopodia} being considered for listing with the Endangered Species Act in the US \cite{lowry_endangered_2023}, and under review by COSEWIC in Canada \cite{heady_roadmap_2022}, the need for informed management actions is high.

An aspect of these dynamics that are not explored in our model is the possibility of resistance to the disease evolving in the population \cite{Jiao2021-uw}. Populations may evolve an increased recovery rate where recovery is not rare, and the recovery dynamics cannot be ignored \cite{Hollanders2023-gg}. In addition, dispersal of individuals with higher pathogen resistance could promote the persistence of the population in highly infectious patches through an evolutionary rescue effect rather than only through an ecological rescue effect \cite{Jaffe2019-ui}. Evolutionary rescue is one of the primary mechanisms to promote host-pathogen coexistence and can be further explored with an eco-evolutionary model \cite{DiRenzo2018-mc}.

In addition to species adaptation, our model considers a simple spatial structure where the metapopulation follows source-sink dynamics. In reality, species may be distributed in various spatial configurations, with patches of different quality regarding transmission rate. In these cases, increased interaction between patches (either through a higher dispersal of juveniles or patch connectedness) may lead to higher mortality, but endemic equilibria should still exist \cite{Jiao2020-de}. In addition, some patches may act as refugia where the disease is not present, or transmission is low enough that a disease-free equilibrium is achievable \cite{Heard2015-jl}. These previous empirical and theoretical studies suggest that our results on species persistence would still be valid for more complex spatial topologies.

In conclusion, our analysis showed the conditions for a sink host population to persist in the presence of an unrecoverable disease. We focus our numerical exploration on juveniles' maturation rate and dispersal rate as these were inspired by the specific case of \textit{Pycnopodia} management. A possible extension to generalize the results of this model is a two-patch model where bidirectional dispersal (potentially asymmetric) is possible.

\backmatter





\bmhead{Acknowledgements}

The authors extend their gratitude for the support from the Research Center in Pure and Applied Mathematics and the Department of Mathematics at the University of Costa Rica.

\section*{Declarations}

\begin{itemize}
    \item \textbf{Conflict of interest} The authors declare no conflict of interest.
    \item \textbf{Code availability} The source code for the numerical analysis performed in this paper can be found in \href{https://github.com/JimmyCalvoMonge/pycnopodia}{https://github.com/JimmyCalvoMonge/pycnopodia}.
\end{itemize}

\noindent








\begin{appendices}

\section{First patch disease free equilibrium Jacobian calculation}\label{sec:first_patch_dfe_jacobian}

Summing the fourth row to the third and the second row to the first equals

\begin{align*}
\det  \cdot \begin{bmatrix}
        - \frac{prA_{S,1}}{k} - (\alpha+\mu_S) -\lambda & - \frac{prA_{S,1}}{k} - \mu_I -\lambda &  - \frac{2prA_{S,1}}{k} & - \frac{prA_{S,1}}{k} \\
        0 & \beta_1 \frac{J_{S,1}}{N_1} - \mu_I -\lambda & 0 & \beta_1 \frac{J_{S,1}}{N_1} \\
        \alpha & 0 & - \mu_S- \lambda & - \mu_I -\lambda \\
        0 & \beta_1 \frac{A_{S,1}}{N_1} & 0 & \beta_1 \frac{A_{S,1}}{N_1} - \mu_I -\lambda
    \end{bmatrix}.
\end{align*}

Transposing this matrix, we must compute

\begin{align*}
    & \det  \begin{bmatrix}
        - \frac{prA_{S,1}}{k} - (\alpha+\mu_S) -\lambda & 0 & \alpha & 0  \\
        - \frac{prA_{S,1}}{k} - \mu_I -\lambda & \beta_1 \frac{J_{S,1}}{N_1} - \mu_I -\lambda & 0 & \beta_1 \frac{A_{S,1}}{N_1} \\
        - \frac{2prA_{S,1}}{k}  & 0  & - \mu_S- \lambda & 0 \\
        - \frac{prA_{S,1}}{k} & \beta_1 \frac{J_{S,1}}{N_1}& - \mu_I -\lambda  & \beta_1 \frac{A_{S,1}}{N_1} - \mu_I -\lambda
    \end{bmatrix} \\\\
    & = \left( - \frac{prA_{S,1}}{k} - (\alpha+\mu_S) -\lambda \right) \cdot \underbrace{\det  \begin{bmatrix}
        \beta_1 \frac{J_{S,1}}{N_1} - \mu_I -\lambda & 0 & \beta_1 \frac{A_{S,1}}{N_1} \\
        0  & - \mu_S- \lambda & 0 \\
        \beta_1 \frac{J_{S,1}}{N_1}& - \mu_I -\lambda  & \beta_1 \frac{A_{S,1}}{N_1} - \mu_I -\lambda
    \end{bmatrix}}_{(A)}\\
    & + \alpha \cdot \underbrace{\det \begin{bmatrix}
        - \frac{prA_{S,1}}{k} - \mu_I -\lambda & \beta_1 \frac{J_{S,1}}{N_1} - \mu_I -\lambda  & \beta_1 \frac{A_{S,1}}{N_1} \\
        - \frac{2prA_{S,1}}{k}  & 0  &  0 \\
        - \frac{prA_{S,1}}{k} & \beta_1 \frac{J_{S,1}}{N_1}& \beta_1 \frac{A_{S,1}}{N_1} - \mu_I -\lambda
    \end{bmatrix}}_{(B)}
\end{align*}

To compute determinant $(A)$, we employ the second row.

\begin{align*}
    (A) &= - (\mu_S + \lambda) \left( \left(\beta_1 \frac{J_{S,1}}{N_1} - \mu_I -\lambda\right)\cdot \left(\beta_1 \frac{A_{S,1}}{N_1} - \mu_I -\lambda\right) - \beta_1 \frac{J_{S,1}}{N_1}\cdot \beta_1 \frac{A_{S,1}}{N_1} \right) \\
    &= - (\mu_S + \lambda)\left( - \frac{\beta_1}{N_1}(\mu_I+ \lambda)(A_{S,1}+J_{S,1}) + (\mu_I+ \lambda)^2\right)
\end{align*}

Likewise, for computing the determinant $(B)$, we use the second row. 
\begin{align*}
    (B) & = \frac{2prA_{S,1}}{k}\left( \left(\beta_1 \frac{J_{S,1}}{N_1} - \mu_I -\lambda\right)\cdot \left(\beta_1 \frac{A_{S,1}}{N_1} - \mu_I -\lambda\right) - \beta_1 \frac{J_{S,1}}{N_1}\cdot \beta_1 \frac{A_{S,1}}{N_1} \right) \\
    &= \frac{2prA_{S,1}}{k}\left( - \frac{\beta_1}{N_1}(\mu_I+ \lambda)(A_{S,1}+J_{S,1}) + (\mu_I+ \lambda)^2\right)
\end{align*}

Therefore, the final determinant equals

\begin{align*}
    \left[- \left( - \frac{prA_{S,1}}{k} - (\alpha+\mu_S) -\lambda \right) \cdot (\mu_S + \lambda) + \alpha \cdot \frac{2prA_{S,1}}{k} \right] \left[ - \frac{\beta_1}{N_1}(\mu_I+ \lambda)(A_{S,1}+J_{S,1}) + (\mu_I+ \lambda)^2\right].
\end{align*}

\section{First patch endemic equilibrium calculation} \label{sec:first_patch_ee_calculation}
Using $I_1:= J_{I,1}+A_{I,1}$ we have that
\begin{align*}
    \dot{I_1} = \dot{J_{I,1}} + \dot{A_{I,1}} &= \beta_1J_{S,1}\frac{(J_{I,1}+A_{I,1})}{N_1}-\mu_IJ_{I,1} + \beta_1A_{S,1}\frac{(J_{I,1}+A_{I,1})}{N_1}-\mu_IA_{I,1} \\
    &=\beta_1(J_{S,1}+A_{S,1})\frac{I_1}{N_1}-\mu_I I_1 \\
\end{align*}
We can then consider the sub-model for the first patch,
\begin{equation}\label{eq:baseModel2}
    \begin{split}
        \dot{J_{S,1}}=&pr\left(1- \frac{N_1}{k}\right)A_{S,1}-\beta_1J_{S,1}\frac{I_1}{N_1}-\alpha J_{S,1}-\mu_SJ_{S,1}\\
        \dot{A_{S,1}}=&\alpha J_{S,1}-\beta_1A_{S,1}\frac{I_1}{N_1}-\mu_SA_{S,1}\\
        \dot{I_1}=&\beta_1(J_{S,1}+A_{S,1})\frac{I_1}{N_1}-\mu_I I_1.
    \end{split}
\end{equation}

We want to find an equilibrium point in this sub-model, that is, a point in which the following equations are satisfied,

\begin{equation}\label{eq:subModeleq1}
    0=pr\left(1- \frac{N_1}{k}\right)A_{S,1}-\beta_1J_{S,1}\frac{I_1}{N_1}-\alpha J_{S,1}-\mu_SJ_{S,1},
\end{equation}
\begin{equation}\label{eq:subModeleq2}
    0=\alpha J_{S,1}-\beta_1A_{S,1}\frac{I_1}{N_1}-\mu_SA_{S,1},
\end{equation}
\begin{equation}\label{eq:subModeleq3}
    0=\beta_1(J_{S,1}+A_{S,1})\frac{I_1}{N_1}-\mu_I I.
\end{equation}

In particular, we are interested in finding an endemic equilibrium, which is an equilibrium in which $I_1 \neq 0$. Using Equation (\ref{eq:subModeleq3}) and canceling $I_1$, we obtain that
\begin{equation}\label{eq:subModeleq4}
    N_1 = \frac{\beta_1}{\mu_I}(J_{S,1} + A_{S,1}).
\end{equation}

Furthermore, using (\ref{eq:subModeleq4}) we get

\begin{align}\label{eq:subModeleq5}
    I_1 :&= N_1 - A_{S,1} - J_{S,1} \nonumber \\
    &= \frac{\beta_1}{\mu_I}(J_{S,1} + A_{S,1}) - (J_{S,1} + A_{S,1}) \nonumber \\
    &= \left[\frac{\beta_1}{\mu_I} -1 \right](J_{S,1} + A_{S,1}) \nonumber \\
    &= \frac{\beta_1 - \mu_I}{\mu_I}(J_{S,1} + A_{S,1})
\end{align}

As $I \neq 0$ we must have $\beta_1 \neq \mu_I$. If we multiply Equation (\ref{eq:subModeleq2}) by $N_1$ and use (\ref{eq:subModeleq4}) y (\ref{eq:subModeleq5}), we obtain 

\begin{align*}
    0&=\alpha J_{S,1}N_1-\beta_1A_{S,1}I_1-\mu_SA_{S,1}N_1 \Rightarrow \\
    0&=\alpha J_{S,1} \cdot \frac{\beta_1}{\mu_I}(J_{S,1} + A_{S,1}) - \beta_1 A_{S,1} \cdot \frac{\beta_1 - \mu_I}{\mu_I}(J_{S,1} + A_{S,1}) -\mu_SA_{S,1} \cdot \frac{\beta_1}{\mu_I}(J_{S,1} + A_{S,1}) \Rightarrow \\
    0&=(J_{S,1} + A_{S,1}) \left[ \alpha J_{S,1} \cdot \frac{\beta_1}{\mu_I} - \beta_1 A_{S,1} \cdot \frac{\beta_1 - \mu_I}{\mu_I} -\mu_SA_{S,1} \cdot \frac{\beta_1}{\mu_I}\right]\Rightarrow \left(\text{clearing } \frac{\beta_1}{\mu_I}\right)\\
    0&=(J_{S,1} + A_{S,1}) \biggl( \alpha J_{S,1} - A_{S,1} (\beta_1 - \mu_I  +\mu_S) \biggr).
\end{align*}

Note that $J_{S,1}+A_{S,1} \neq 0$ because otherwise (\ref{eq:subModeleq3}) would imply $I_1=0$. Then it follows that $\alpha J_{S,1} =  A_{S,1} (\beta_1 - \mu_I  +\mu_S)$. Moreover, if we assume $\beta_1 - \mu_I  +\mu_S \neq 0$ we arrive at the equality

\begin{equation}\label{eq:subModeleq6}
    \frac{\alpha}{\beta_1 - \mu_I  +\mu_S}J_{S,1} = A_{S,1}.
\end{equation}

Call $\lambda_1 := \frac{\alpha}{\beta_1 - \mu_I  +\mu_S} = \frac{\alpha}{\mu_I \left(R_{0,1} + \frac{\mu_S}{\mu_I} - 1 \right)}$ and note that
\begin{align*}
    N_1 = A_{S,1} + J_{S,1} + I_1 = \lambda_1 J_{S,I} + J_{S,1} + I_1 = (\lambda_1+1)J_{S,1} + I_1.
\end{align*}

Modifying (\ref{eq:subModeleq5}) we get 

\begin{align*}
    I_1 = \frac{\beta_1 - \mu_I}{\mu_I}(J_{S,1} + A_{S,1}) = \frac{\beta_1 - \mu_I}{\mu_I}(\lambda_1+1)J_{S,1} \Rightarrow \frac{\mu_I}{\beta_1 - \mu_I} I_1 = (\lambda_1+1)J_{S,1} 
\end{align*}

meaning that

\begin{equation}\label{eq:subModeleq7}
    N_1 = (\lambda_1+1)J_{S,1} + I_1 = \frac{\mu_I}{\beta_1 - \mu_I} I_1 + I_1 = \frac{\beta_1}{\beta_1 - \mu_I}I_1.
\end{equation}

We turn to the first equation of the model, Equation (\ref{eq:subModeleq1}), and substitute Equation (\ref{eq:subModeleq6}) to obtain

\begin{align*}
    0=pr\left(1- \frac{N_1}{k}\right)\lambda_1 J_{S,1} -\beta_1J_{S,1}\frac{I_1}{N_1}-\alpha J_{S,1}-\mu_SJ_{S,1}.
\end{align*}

Note that $J_{S,1}=0$ would imply $A_{S,1}$ because of (\ref{eq:subModeleq6}), which implies $J_{S,1}+A_{S,1}=0$, an impossibility when assuming $I_1 \neq 0$ as we commented above. Therefore we can cancel $J_{S,1}$ in this last equation to get

\begin{align*}
    0=pr\lambda_1\left(1- \frac{N_1}{k}\right) -\beta_1\frac{I_1}{N_1}-(\alpha + \mu_S).
\end{align*}

If we multiply by $N_1$, we get

\begin{align*}
    0=pr\lambda_1\left(1- \frac{N_1}{k}\right)N_1 -\beta_1I-(\alpha + \mu_S)N_1.
\end{align*}

Using (\ref{eq:subModeleq7}) we obtain

\begin{align*}
    0=pr\lambda_1\left(1- \frac{\beta_1}{k(\beta_1 - \mu_I)}I_1\right)\frac{\beta_1}{\beta_1 - \mu_I}I_1 -\beta_1I_1-(\alpha + \mu_S)\frac{\beta_1}{\beta_1 - \mu_I}I_1.
\end{align*}

As $I_1 \neq 0$ we can cancel all $I_1$ values and arrive to 

\begin{align}\label{eq:clearIpre}
    0=pr\lambda_1\left(1- \frac{\beta_1}{k(\beta_1 - \mu_I)}I_1\right)\frac{\beta_1}{\beta_1 - \mu_I} -\beta_1-\frac{\beta_1(\alpha + \mu_S)}{\beta_1 - \mu_I}.
\end{align}

We note that $\frac{\beta_1}{\beta_1 - \mu_I} = \frac{R_{0,1}}{R_{0,1} - 1}$, so we have that

\begin{align}\label{eq:clearIpreR0}
    0=pr\lambda_1\left(1- \frac{R_{0,1}}{k(R_{0,1}-1)}I_1\right)\frac{R_{0,1}}{R_{0,1} - 1}-\mu_I R_{0,1} -\frac{R_{0,1}(\alpha + \mu_S)}{R_{0,1}-1}.
\end{align}

Clearing $I_1$ we get

\begin{equation}\label{eq:subModel8}
    I_1^* = \frac{k\left(R_{0,1}-1\right)}{pr\alpha R_{0,1}}\biggl(pr\alpha - \bigl(\mu_IR_{0,1} + \mu_S - \mu_I  + \alpha\bigr)\bigl(\mu_I R_{0,1} + \mu_S - \mu_I\bigr)\biggr).
\end{equation}

Further normalizing with respect to $\mu_I$, we obtain that

\begin{equation}\label{eq:subModel8_2}
    I_1^* = \frac{k\left(R_{0,1}-1\right)}{r R_p R_{\alpha} R_{0,1}}\biggl(rR_pR_{\alpha} - (R_{0,1} + R_s + R_{\alpha} -1)(R_{0,1} + R_s -1)\biggr),
\end{equation}
where
\begin{equation}\label{eq:R_definitions}
    R_p := \frac{p}{\mu_I}, \quad R_{\alpha} := \frac{\alpha}{\mu_I}, \quad R_s = \frac{\mu_S}{\mu_I}.
\end{equation}

We can obtain the values of $J_{S,1}$ and $A_{S,1}$ at the equilibrium with the previous equations. These are given by

\begin{align}\label{def:vals_a_j_1_equilibrium}
    J_{S,1}^* &=  \frac{\mu_I}{(\beta_1 - \mu_I)(\lambda_1+1)}I_1^* = \frac{1}{(R_{0,1}-1)(\lambda_1+1)}I_1^*,\nonumber \\
    A_{S,1}^* &= \frac{\mu_I\lambda_1}{(\beta_1 - \mu_I)(\lambda_1+1)}I_1^* = \frac{\lambda_1}{(R_{0,1}-1)(\lambda_1+1)}I_1^*.
\end{align}

Finally, we can return to the original model and find the values of $J_{I,1}$ and $A_{I,1}$ at this equilibrium. Using $\dot{J_{S,1}}=0$ in the second Equation (\ref{eq:baseModel}) and plugging the values of $I_1^*, J_{S,1}^*$ and $N_1^* = R_{\beta_1}I_1^*$ we get 
\begin{align*}
    0 &= \beta_1 \cdot \frac{\mu_IR_{\beta_1}}{\beta_1(\lambda_1+1)}I_1^* \cdot \frac{I_1^*}{R_{\beta_1} I_1^*} - \mu_I J_{I,1}^* \Rightarrow J_{I,1}^* = \frac{1}{\lambda_1+1}I_1^* \\
    & \Rightarrow A_{I,1}^* = \frac{\lambda_1}{\lambda_1+1}I_1^*.
\end{align*}

\section{Second patch disease free equilibrium Jacobian calculation}\label{sec:second_patch_dfe_jacobian}

By adding the term $-\frac{(1-p)r}{k}A_{S,1}$ to each entry of the first row and transposing this matrix, we must compute

\begin{align*}
    & \det  \begin{bmatrix}
        - \frac{prA_{S,2}}{k} - (\alpha+\mu_S) -\lambda -\frac{(1-p)r}{k}A_{S,1} & 0 & \alpha & 0  \\
        - \frac{prA_{S,2}}{k} - \mu_I -\lambda -\frac{(1-p)r}{k}A_{S,1} & \beta_2 \frac{J_{S,2}}{N_2} - \mu_I -\lambda & 0 & \beta_2 \frac{A_{S,2}}{N_2} \\
        - \frac{2prA_{S,2}}{k}  -\frac{(1-p)r}{k}A_{S,1}& 0  & - \mu_S- \lambda & 0 \\
        - \frac{prA_{S,2}}{k} -\frac{(1-p)r}{k}A_{S,1}& \beta_2 \frac{J_{S,2}}{N_2}& - \mu_I -\lambda  & \beta_2 \frac{A_{S,2}}{N_2} - \mu_I -\lambda
    \end{bmatrix} \\\\
    & = \left( - \frac{prA_{S,2}}{k} - (\alpha+\mu_S) -\lambda -\frac{(1-p)r}{k}A_{S,1} \right) \cdot \underbrace{\det  \begin{bmatrix}
        \beta_2 \frac{J_{S,2}}{N_2} - \mu_I -\lambda & 0 & \beta_2 \frac{A_{S,2}}{N_2} \\
        0  & - \mu_S- \lambda & 0 \\
        \beta_2 \frac{J_{S,2}}{N_2}& - \mu_I -\lambda  & \beta_2 \frac{A_{S,2}}{N_2} - \mu_I -\lambda
    \end{bmatrix}}_{(A)}\\
    & + \alpha \cdot \underbrace{\det \begin{bmatrix}
        - \frac{prA_{S,2}}{k} - \mu_I -\lambda -\frac{(1-p)r}{k}A_{S,1}& \beta_2 \frac{J_{S,2}}{N_2} - \mu_I -\lambda  & \beta_2 \frac{A_{S,2}}{N_2} \\
        - \frac{2prA_{S,2}}{k} -\frac{(1-p)r}{k}A_{S,1} & 0  &  0 \\
        - \frac{prA_{S,2}}{k}-\frac{(1-p)r}{k}A_{S,1} & \beta_2 \frac{J_{S,2}}{N_2}& \beta_2 \frac{A_{S,2}}{N_2} - \mu_I -\lambda
    \end{bmatrix}}_{(B)}
\end{align*}

To compute determinant $(A)$, we employ the second row.

\begin{align*}
    (A) &= - (\mu_S + \lambda) \left( \left(\beta_2 \frac{J_{S,2}}{N_2} - \mu_I -\lambda\right)\cdot \left(\beta_2 \frac{A_{S,2}}{N_2} - \mu_I -\lambda\right) - \beta_2 \frac{J_{S,2}}{N_2}\cdot \beta_2 \frac{A_{S,2}}{N_2} \right) \\
    &= - (\mu_S + \lambda)\left( - \frac{\beta_2}{N_2}(\mu_I+ \lambda)(A_{S,2}+J_{S,2}) + (\mu_I+ \lambda)^2\right)
\end{align*}

Likewise, for computing the determinant $(B)$, we use the second row. 
\begin{align*}
    (B) & = \left(\frac{2prA_{S,2}}{k} +\frac{(1-p)r}{k}A_{S,1}\right) \left( \left(\beta_2 \frac{J_{S,2}}{N_2} - \mu_I -\lambda\right)\cdot \left(\beta_2 \frac{A_{S,2}}{N_2} - \mu_I -\lambda\right) - \beta_2 \frac{J_{S,2}}{N_2}\cdot \beta_2 \frac{A_{S,2}}{N_2} \right) \\
    &= \left(\frac{2prA_{S,2}}{k} + \frac{(1-p)r}{k}A_{S,1}\right)\left( - \frac{\beta_2}{N_2}(\mu_I+ \lambda)(A_{S,2}+J_{S,2}) + (\mu_I+ \lambda)^2\right)
\end{align*}

We note that by grouping the two terms $\frac{(1-p)r}{k}A_{S,1}$ cancel out, and the final form of the determinant comes to be

\begin{align*}
    \left[- \left( - \frac{prA_{S,2}}{k} - (\alpha+\mu_S) -\lambda \right) \cdot (\mu_S + \lambda) + \alpha \cdot \frac{2prA_{S,2}}{k} \right] \left[ - \frac{\beta_2}{N_2}(\mu_I+ \lambda)(A_{S,2}+J_{S,2}) + (\mu_I+ \lambda)^2\right].
\end{align*}
Just as in the case of the first patch.




\end{appendices}


\bibliography{references}

\end{document}